\documentclass{elsart}

\newcommand{\bmath}[1]{\mbox{\boldmath $#1$}}

\usepackage{graphicx,amssymb,amsmath}
\usepackage{cite}
\begin{document}
\begin{frontmatter}
\title{A precision determination of the mass of the $\eta$ meson}
\author{The GEM Collaboration:}
\author[a]{M.~Abdel-Bary},
\author[d]{A.~Budzanowski},
\author[i]{A.~Chatterjee},
\author[g]{J.~Ernst},
\author[a,c]{P.~Hawranek},
\author[g]{R.~Jahn},
\author[i]{V.~Jha},
\author[a]{K.~Kilian},
\author[d]{S.~Kliczewski},
\author[n]{D.~Kirillov},
\author[f]{D.~Kolev},
\author[m]{M.~Kravcikova},
\author[e]{T.~Kutsarova},
\author[c]{M.~Lesiak},
\author[j]{J.~Lieb},
\author[a]{H.~Machner}\corauth[cor]{Corresponding author}\ead{h.machner@fz-juelich.de},
\author[c]{A.~Magiera},
\author[a]{R.~Maier},
\author[l]{G.~Martinska},
\author[k]{S.~Nedev},
\author[n]{N.~Piskunov},
\author[a]{D.~Prasuhn},
\author[a]{D.~Proti\'c},
\author[a]{P.~von Rossen},
\author[a,i]{B.~J.~Roy},
\author[n]{I.~Sitnik},
\author[d,g]{R.~Siudak},
\author[c]{M.~Smiechowicz},
\author[a]{H.~J.~Stein},
\author[f]{R.~Tsenov},
\author[a,l]{M.~Ulicny},
\author[a,g]{J.~Urban},
\author[a,f]{G.~Vankova},
\author[o]{C.~Wilkin}
\address[a]{Institut f\"{u}r Kernphysik, Forschungszentrum J\"{u}lich, J\"{u}lich, Germany}
\address[d]{Institute of Nuclear Physics, Polish Academy of Sciences, Krakow, Poland}
\address[i]{Nuclear Physics Division, BARC, Bombay, India}
\address[g]{Helmholtz-Institut f\"{u}r Strahlen- und Kernphysik der Universit\"{a}t Bonn, Bonn, Germany}
\address[c]{Institute of Physics, Jagellonian University, Krakow, Poland}
\address[n]{Laboratory for High Energies, JINR Dubna, Russia}
\address[f]{Physics Faculty, University of Sofia, Sofia, Bulgaria}
\address[m]{Technical University, Kosice, Kosice, Slovakia}
\address[e]{Institute of Nuclear Physics and Nuclear Energy, Sofia, Bulgaria}
\address[j]{Physics Department, George Mason University, Fairfax, Virginia, USA}
\address[l]{P.~J.~Safarik University, Kosice, Slovakia}
\address[k]{University of Chemical Technology and Metallurgy, Sofia, Bulgaria}
\address[o]{Department of Physics \& Astronomy, UCL, London, U.K.}
\begin{abstract}%
Several processes of meson production in proton-deuteron collisions
have been measured simultaneously using a calibrated magnetic
spectrograph. Among these processes, the $\eta$ meson is seen
clearly as a sharp missing-mass peak on a slowly varying background
in the $p+d\to\,^3\textrm{He}\,+X$ reaction. Knowing the kinematics
of the other reactions with well determined masses, it is possible
to deduce a precise mass for the $\eta$ meson. The final result,
$m(\eta)=547.311\pm 0.028\,\textrm{(stat)} \pm
0.032\,\textrm{(syst)~MeV/c}^2$, is significantly lower than that
found by the recent NA48 measurement, though it is consistent with
values obtained in earlier counter experiments.
\end{abstract}
\end{frontmatter}
Compared to other light mesons, the mass of the $\eta$ is
surprisingly poorly known. Though the Particle Data Group (PDG)
quote a value of $m_{\eta}=547.75\pm 0.12~\textrm{MeV/c}^2$ in their
2004 review~\cite{PDG04}, this error hides differences of up to
0.7~MeV/c$^2$ between the results of some of the modern counter
experiments quoted. The PDG average is in fact dominated by the
result of the CERN NA48 experiment,
$m_{\eta}=547.843\pm0.051~\textrm{MeV/c}^2$, which is based upon the
study of the kinematics of the six photons from the $3\pi^0$ decay
of 110~GeV $\eta$-mesons~\cite{Lai02}. In the other experiments
employing electronic detectors, which typically suggest a mass
$\approx 0.5~\textrm{MeV/c}^2$ lighter, the $\eta$ was produced much
closer to threshold and its mass primarily determined through a
missing-mass technique where, unlike the NA48 experiment, precise
knowledge of the beam momentum plays an essential part. In the
Rutherford Laboratory experiment the momentum of the pion beam in
the $\pi^-+p\to n+\eta$ reaction was fixed macroscopically using the
floating wire technique~\cite{Duane74}. In the measurement making
use of the photoproduction reaction $\gamma+p\to p+\eta$, the energy
of the electrons that were the source of the bremsstrahlung photons
was fixed to a relative precision of $2\times 10^{-4}$ by measuring
the distance of the beam paths in the third race track microtron of
the MAMI accelerator~\cite{Krusche95}. In the Saclay SATURNE
experiment a high resolution, but small acceptance, spectrometer was
used and, through an ingenious series of measurements on different
nuclear reactions, the beam energy and spectrograph settings were
both calibrated. The value of the $\eta$ mass was then extracted
from the missing-mass peak in the $p+d\to\,^3\textrm{He}\,+X$
reaction~\cite{Plouin92}.

In an attempt to clarify the situation, we have performed an
experiment at COSY in J\"{u}lich specifically designed to determine the
$\eta$-mass with high precision. The methodology is very similar in
spirit to that used at SATURNE~\cite{Plouin92} in that several
reactions were measured, thus allowing one to calibrate the
accelerator beam and the detector and hence to measure the
$\eta$-mass with potentially a very small systematic error. The
crucial difference from SATURNE is that the spectrograph (Big Karl)
that we have used has a large acceptance. Therefore all the
reactions could be studied simultaneously.

Following ideas already developed in Ref.~\cite{1930}, we have
measured simultaneously the following three reactions:
\begin{eqnarray}
p+d&\to &\pi^++ t \label{eqn:reaction_pit} \\
p+d&\to &t+\pi^+  \label{eqn:reaction_t} \\
p+d&\to &^3\textrm{He}\, +\eta ,\label{eqn:reaction_He}
\end{eqnarray}
where in each case it was the first particle on the right that was
measured.

\begin{figure}\centering
\includegraphics[width=8 cm]{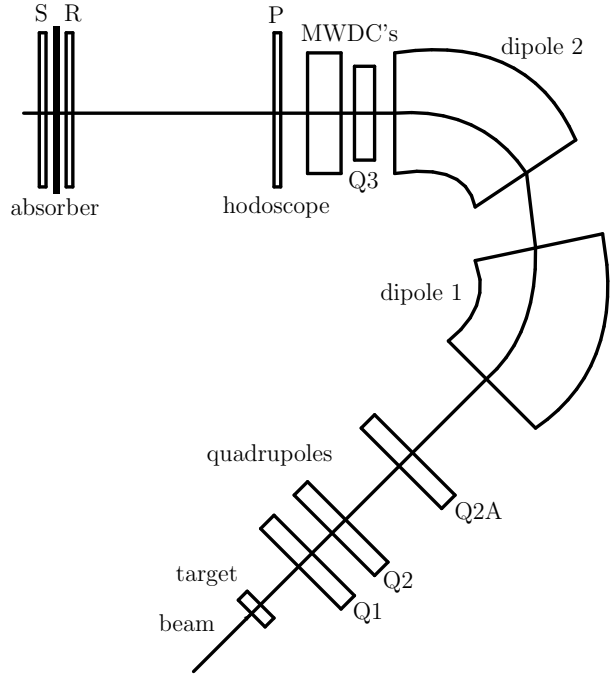}
\caption{Sketch of the Big Karl magnetic spectrograph and the
focal plane detector arrangement.} \label{BIG_KARL}
\end{figure}

Charged particles were detected with the help of Big Karl, a
focussing 3Q2DQ magnetic spectrograph whose principal elements are
indicated in Fig.~\ref{BIG_KARL}. It should be noted that the final
quadrupole magnet $Q3$ was not actually used in this study. We
define the beam to be incident in the $z$-direction, with the
$y$-direction being vertical, and the $x$-direction horizontal and
perpendicular to the beam. The optics with respect to the horizontal
and vertical motions are almost decoupled in Big
Karl~\cite{Drochner98}. In the horizontal direction the spectrograph
has a point-to-point imaging from the target to the focal plane with
dispersion whereas, in the vertical direction, it operates in the
point-to-parallel mode.

Tracks were measured with two packs of multiwire drift chambers
(MWDCs). Each pack consists of six layers, allowing a precise
determination of the position of a charged particle. The drift time
measurement was started by a signal from the hodoscope layer P and
an individual drift time calibration was performed for each particle
type. Signals from hodoscope layers P and R, approximately 3.5$\:$m
apart, were used for a time-of-flight measurement. Together with
specific energy loss in the scintillators and the momentum vector,
this allowed particle identification and hence a determination of
the energy of the particle. It is important to note that at a beam
momentum around 1640~MeV/c all three reactions can be observed
simultaneously with a single setting of the Big Karl magnetic
fields. The first two reactions were used to calibrate the beam and
the spectrograph with the third determining $m_{\eta}$.

The precision of a missing mass measurement depends on the accuracy
with which the four-momentum vectors of the incident particles in
the entrance channel and of the detected particle in the exit
channel are known. In order to define well the reaction vertex, a
liquid deuterium target only 2~mm thick was employed
\cite{Jaeckle94}. The Mylar windows were only 1~$\mu$m thick, thus
making background reactions in the window material negligible. The
target was operated at a temperature of 18.7 K which can lead to
freezing out of residual gas on the windows. The target was
therefore cleaned by warming it up periodically. The proton beam at
COSY was electron-cooled at injection energy and then stochastically
extracted, which resulted in the following beam properties (uncooled
beam properties in brackets): $\Delta p/p_0 = 7.5\times 10^{-5}\
(3.2\times 10^{-4})$, $\Delta p_x/p_0=0.9\ (1.8)$~mrad, $\Delta
p_y/p_0=0.8\ (5.8)$~mrad. It can be seen that electron cooling gave
an important improvement for this experiment. Another benefit from
the cooling was the reduction in the beam halo and hence in the
associated background.

We start by discussing the principles of a momentum measurement with
the magnetic spectrograph. At any specified position in the system,
an arbitrary charged particle is represented by a column vector
$\bmath{V}$, whose components are the positions, angles, and
momentum of the particle with respect to the reference trajectory,
which is chosen to be the $z$-axis. This vector then reads:
\begin{equation}
\bmath{V}=\left(
\begin{array}{l}
x\\
x'\\
y\\
y'\\
l\\
\delta
\end{array}
\right)
\end{equation}
where we have used the following definitions:
\begin{verse}
$x $ is the horizontal displacement of an arbitrary ray (or particle
track) with respect to the assumed central trajectory.\\
$x' $ is the tangent of the angle that this ray makes in the
horizontal plane with respect to the assumed central trajectory.\\
$y $ is the vertical displacement of the ray with respect to the
assumed central trajectory.\\
$ y' $ is the tangent of the vertical angle of the ray with
respect to the assumed central trajectory.\\
$l$ is the path length difference between the actual ray and the
central trajectory.\\
$\delta=\Delta p/p$ is the fractional momentum deviation of the ray
from the assumed central trajectory which corresponds to the assumed
Big Karl central momentum $p_{BK}$.
\end{verse}
For any two different positions in the overall system, such as the
target ($t$) and the focal plane ($f$), the corresponding vectors
are connected through the transport matrix $\bmath{R}$
\begin{equation}
\bmath{V}_{\!t}=\bmath{R}\, \bmath{V}_{\!f}\,.
\end{equation}

The matrix elements were not necessary constant and, where needed,
they were expanded in powers of $\delta$. The most important
parameter in the transport matrix for the reconstruction of the
momentum in the present case is the element $R_{16}$, which is the
dispersion. Some of the other elements are either zero or small and
can be neglected, whereas others, including $R_{16}$, must be
determined using data from calibration measurements, which will be
discussed now.

A low intensity 793~MeV/c proton beam and empty target were used to
investigate the dependence of $\bmath{R}$ on $\delta$. For this
purpose the central momentum $p_{BK}$ was changed and the proton
tracks reconstructed in the drift chambers. Before each such
measurement the magnetic fields in the dipoles were set and measured
with nuclear magnetic resonance probes; the differences between the
predicted and measured values of the field was of order $10^{-5}$.
We measured in this way twice at 17 values of the central momentum.

The next calibration was based upon detecting the deuterons from the
$p+p\rightarrow d+\pi^+$ reaction, also at 793~MeV/c. At this
momentum the spectrograph has full acceptance for this reaction and
this momentum is close to 804.4 MeV/c, the central spectrograph
momentum, where all three reactions (\ref{eqn:reaction_pit}),
(\ref{eqn:reaction_t}), and (\ref{eqn:reaction_He}) fit into the
acceptance. In the time-of-flight part of the set-up shown in
Fig.~\ref{BIG_KARL} we used an additional scintillator layer, S, in
the veto mode. A 5~cm aluminum absorber was placed between this and
the R layer. This thickness was sufficient to stop deuterons, but
not protons with the same momentum. This reduced the background
originating from the direct beam protons that have a momentum close
to those of detected deuterons.

Finally we studied the $p+p\to\pi^++d$ reaction at a beam momentum
of 1642~MeV/c. The central momentum of the spectrograph was again
varied and the pions measured for 12 different field settings.

From each calibration experiment the values of the possible
parameters were extracted and these were used as start values to fix
all the elements of the transport matrix $R_{ij}$ in one
least-squares fit to all of the calibration measurements.

\begin{figure}
\centering
\includegraphics[width=8 cm]{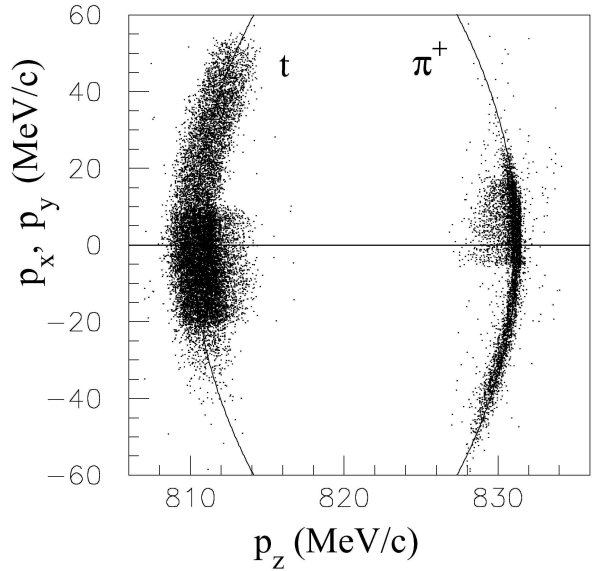}
\caption{Plot of $(p_x,p_z)$ and $(p_y,p_z)$ for coincident events
from reactions (1) and (2). The solid curves give the predictions
for the mean loci for such coincidences. The points in a restricted
range of ordinate values with a larger scatter on the abscissa are
the $(p_x,p_z)$ data points. The different scatter result from the
different projections of a rotational ellipsoid onto the different
planes.} \label{kinem}
\end{figure}

For the production runs measuring reactions
(\ref{eqn:reaction_pit}), (\ref{eqn:reaction_t}) and
(\ref{eqn:reaction_He}) simultaneously, the spectrograph was set to
the nominal momentum $p_{BK}=804.5$ MeV/c. The experiment was
performed in a series of runs that were analyzed separately. This
analysis yielded an unexpected result, indicating a change in some
parameters with time. One possible cause could have been a variation
of the magnetic fields in the dipoles. However, this is contrary to
the very high precision measurements of the fields with NMR probes
and so the variation must have another origin. This was found when
inspecting the target thickness as a function of time.

Although the target was rather thin, corrections were made for the
energy losses of the particles. While for pions this correction is
negligible, and for tritons it is modest, for ${^3\textrm{He}}$
ions it is significant. We then proceeded as follows. The
$804.5$~MeV/c setting was, in a first approximation, assumed to be
exact and constant with time. The beam momentum, target thickness,
and $\eta$-mass were then free parameters to be fit to the data.
In a second step, we checked the approximations made by inverting
the calibration method. The spectrograph was assumed to be at its
nominal value and the missing masses of the particles were
derived. This comparison yields a measure of the precision to
which our method works and give us an estimate of the systematic
errors.

In Fig.~\ref{kinem} coincident events from reactions
(\ref{eqn:reaction_pit}) and (\ref{eqn:reaction_t}) are shown in
terms of their longitudinal and transverse momentum components. The
expected kinematic loci are rotational ellipsoids in three
dimensions.  Projections of their surfaces are shown as curves.

\begin{figure}
\begin{center}
\includegraphics[width=8 cm]{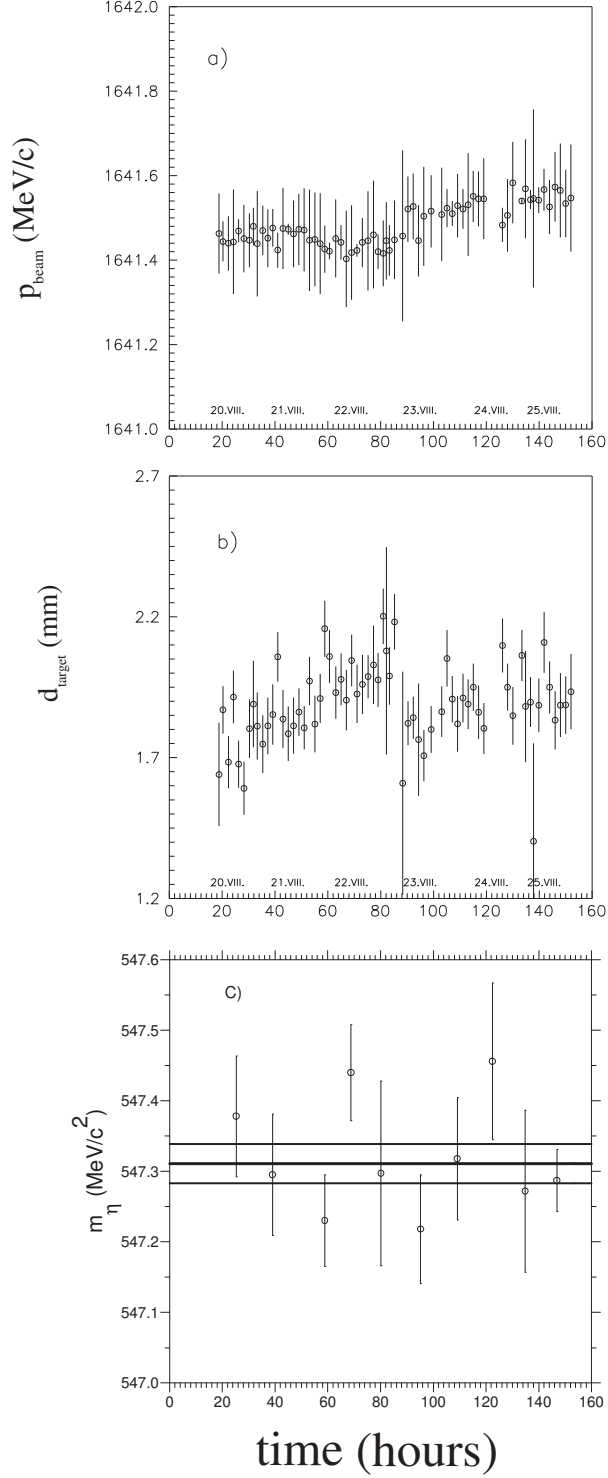}
\caption{a) The reconstructed beam momentum $p_{beam}$ from measured
pions in reaction~(\ref{eqn:reaction_pit}),  b) The target thickness
deduced from the tritons in reaction~(\ref{eqn:reaction_t}), and c)
the $\eta$ mass as functions of the time of measurement. The error
bars shown are purely statistical. The thick solid line denotes the
mean and the two thin lines the $\pm 1\sigma$ error band. }
\label{Beam_target}
\end{center}
\end{figure}

The beam momentum was deduced from the measurement of the pion
four-momenta, which are almost unaffected by the target thickness
(see Fig.~\ref{kinem}). The target thickness was then deduced from
the measurement of the four-momenta of the tritons. The results of
these two measurements are shown in Fig.~\ref{Beam_target} as
functions of the time of measurement. The beam momentum was found to
be quite stable, with a variation from the beginning to the end of
the experiment of only $3\times 10^{-5}$!  However, the target
thickness showed a steady increase with time. Sixty hours after the
beginning of the experiment there was an interruption during which
the target was warmed up and any possible freeze-out on the windows
was removed. It should be noted that the increase of effective
target thickness corresponds to a freeze-out of $\approx 100~\mu$m
frozen air within $\approx 70$ hours. As shown in
Fig.~\ref{Beam_target}b, after the interruption at 88 hours the
effective target thickness may have started to increase once again.

\begin{figure}
\begin{center}
\includegraphics[width=8 cm]{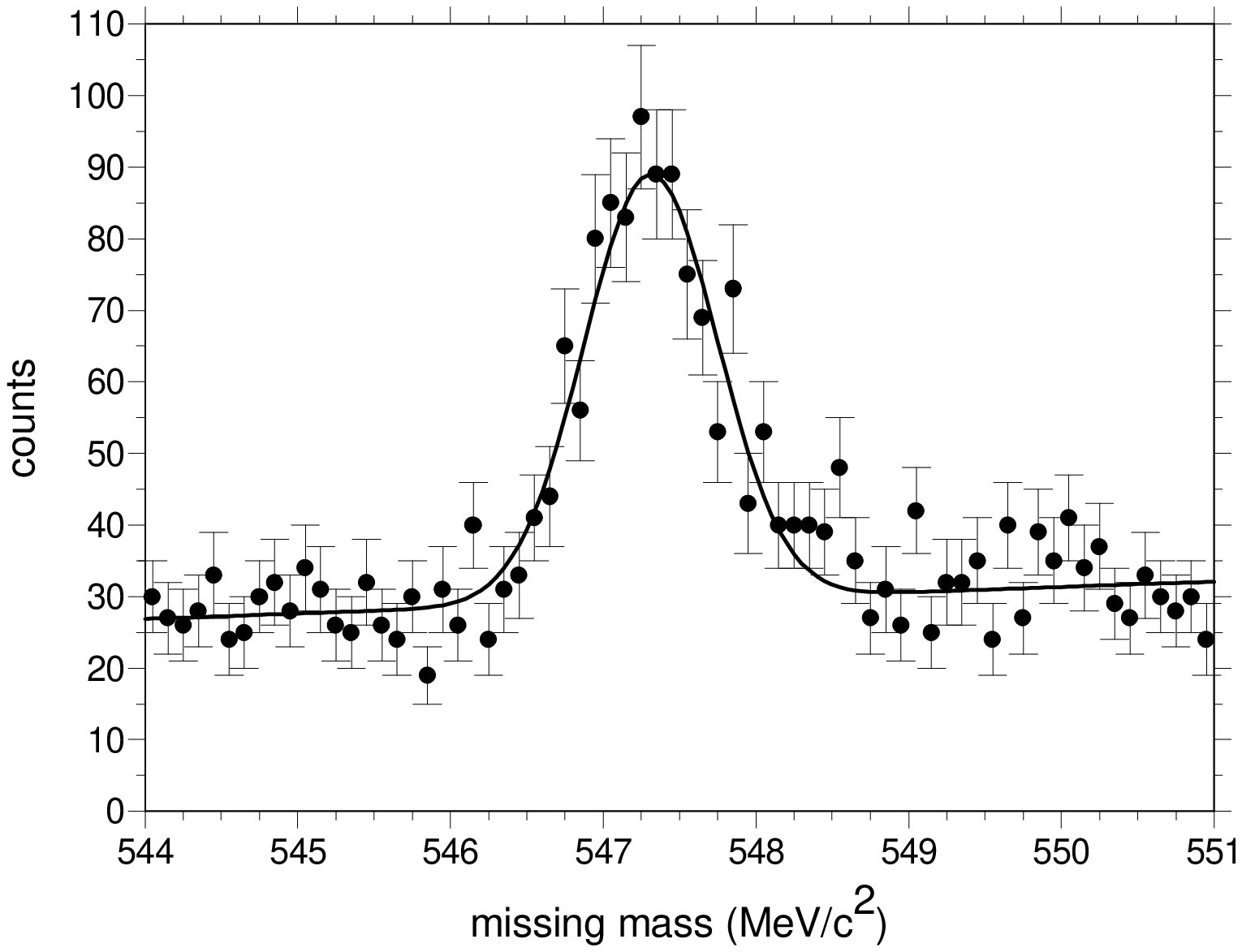}
\caption{The massing mass spectrum from the
$p+d\to\,^3\textrm{He}+X$ reaction. The solid curve shows a fit
with a Gaussian peak superimposed on a linear background.}
\label{eta_missing_mass}
\end{center}
\end{figure}

We are now in a position to extract the value of $m_{\eta}$ from the
missing mass distribution in the $p+d\to\,^3\textrm{He}+X$ reaction
using the four-momenta of the ${^3\textrm{He}}$-ions measured
simultaneously with reactions (\ref{eqn:reaction_pit}) and
(\ref{eqn:reaction_t}).  Our extracted values of the $\eta$ mass are
shown as a function of measuring time in Fig.~\ref{Beam_target}c
where, because of the limited count rate, several runs have been
grouped together. No correlation is visible with the other two
reconstructed observables. Also shown is the mean value and the
uncertainty. The combined missing mass distribution for all events
is shown in Fig.~\ref{eta_missing_mass}, together with a fit in
terms of a Gaussian peak on top of an almost constant linear
background corresponding to multipion production. The width of the
$p+d\to\,^3\textrm{He}+\eta$ peak is in accord with Monte Carlo
simulations of this reaction.

\begin{figure}
\begin{center}
\includegraphics[width=8 cm]{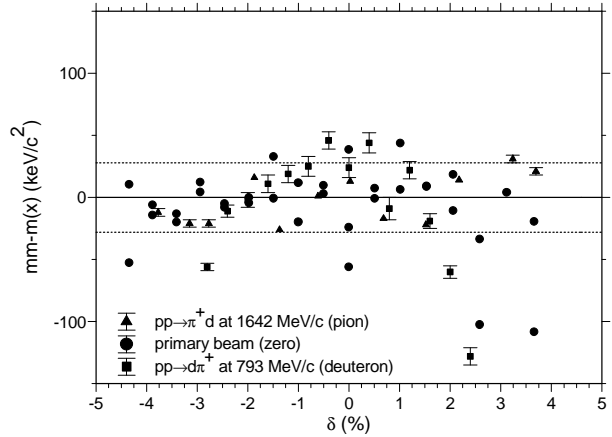}
\caption{The deviation of the measured missing mass $mm$ from the
rest mass $m(x)$ \cite{PDG04} for the particle types $x$ as function
of the deviation of the mean momentum setting $\delta$ of the
spectrograph. The horizontal lines indicate the $1\sigma$ band of
$\pm 28$ keV/c$^2$.} \label{syst_error}
\end{center}
\end{figure}

To get an estimate of some of the systematic errors, we investigated
the influence of the assumption that the mean momentum setting of
the spectrograph is known. For this purpose we applied the deduced
parameters and kept the mass of the measured particle as a variable.
In the case of the direct beam this is of course zero whereas for
$p+p\to d+\pi^+$ at 793~MeV/c it is the $\pi^+$ and for $p+p\to
\pi^++d$ at 1642~MeV/c it is the deuteron. The deviations from the
$p$, $\pi^+$ and $d$ masses are shown in Fig.~\ref{syst_error}.

The mean mass differences are $\pm 20$~keV/c$^2$ for protons, $\pm
32$~keV/c$^2$ for pions, and $\pm 21$~keV/c$^2$ for deuterons. The
average of these, which is one measure of the systematic error, is
$\pm 28$ keV/c$^2$, and this interval is shown in
Fig.~\ref{syst_error}. Now there seems to be a stronger deviation of
the missing mass from the true value for larger positive values of
$\delta$ but it is important to note that the $\eta$-mass was
determined at the position $\delta\approx -2.8\%$ where the
deviation is minimal.

Another systematic error arises from the uncertainty in the liquid
deuterium density depending on the target temperature. This
uncertainty was studied with the help of the codes
GEANT3~\cite{GEANT} and SRIM~\cite{TRIM}, gives only an additional
0.004~MeV/c$^2$ to the systematic error.

The final result of our measurement is
\begin{equation}
m_{\eta}=547.311\pm 0.028\ \textrm{(stat.)} \pm 0.032\
\textrm{(syst.)\ MeV/c}^2\,.
\end{equation}

\begin{figure}
\begin{center}
\includegraphics[width=8cm]{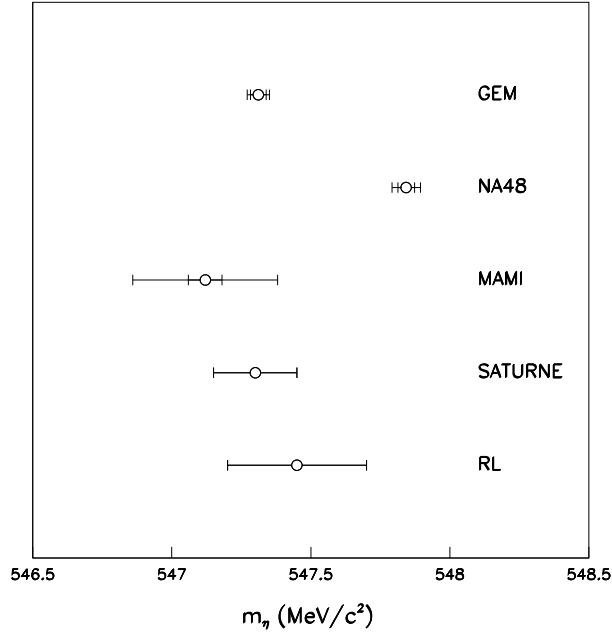}
\caption{The results of the $\eta$-mass measurements, in order of
publication date, taken from the Rutherford Laboratory
(RL)~\cite{Duane74}, SATURNE~\cite{Plouin92}, MAMI~\cite{Krusche95},
NA48~\cite{Lai02}, and GEM. When two error bars are shown, the
smaller is statistical and the larger total.} \label{eta_year}
\end{center}
\end{figure}

Our value of the mass of the $\eta$ meson is compared in
Fig.~\ref{eta_year} with the results of all other measurements
taken into consideration in the current PDG
compilation~\cite{PDG04}. Though significantly smaller than that
reported in the NA48 experiment~\cite{Lai02}, it is in excellent
agreement with the other results. This is very puzzling in that
the NA48 experiment yields an excellent value for the $K^0$ mass,
also through the $3\pi^0$ decay, though the statistics were then
much higher and the systematics not completely identical.\\

The quality of the beam necessary for the success of this work is
due mainly to the efforts of the COSY operator crew. Support by
Internationales B\"uro des BMBF (IND 01/022), SGA, Slovakia
(1/1020/04), and the Forschungszentrum J\"ulich is gratefully
acknowledged.

\end{document}